\pdfoutput=1
\documentclass[twocolumn,
               showpacs,
               preprintnumbers,
               nofootinbib,
               prd,
               superscriptaddress,
               10pt,
               notitlepage,
               aps]{revtex4-2}
               
\usepackage{graphicx,amssymb,amsmath,amsthm,amsfonts,epsfig,mathtools}

\usepackage[utf8]{inputenc}
\usepackage{graphicx}
\usepackage{dcolumn}
\usepackage{bm}
\usepackage{amsmath}
\usepackage{color}
\usepackage[dvipsnames]{xcolor}
\usepackage{hyperref}
\hypersetup{colorlinks=true, citecolor=MidnightBlue,linkcolor=CornflowerBlue, urlcolor=CornflowerBlue, linktocpage=true}
\usepackage{xfrac}
\usepackage{siunitx}
\usepackage{soul}
\newcommand{\beqn}{\begin{eqnarray}}
\newcommand{\eeqn}{\end{eqnarray}}
\newcommand{\beq}{\begin{equation}}
\newcommand{\eeq}{\end{equation}}

\def\d{\mathrm{d}}

\def\gbar{\bar{g}}

\begin{document}

\title{Loss of hyperbolicity and tachyons in generalized Proca theories}

\begin{abstract}
Various groups recently demonstrated that the time evolution of the simplest self-interacting vector fields, those with self-interaction potentials, can break down after a finite duration in what is called loss of hyperbolicity. We establish that this is not an isolated issue, and other generalizations of the Proca theory suffer from the same problem. Specifically, we show that vector field theories with derivative self-interactions have a similar pathology. For this, we derive the effective metric that governs the dynamics, and show that it can change signature during time evolution. We also show that, generalized Proca theories may suffer from tachyonic instabilities as well, which lead to another form of unphysical behavior.
\end{abstract}

\author{K{\i}van\c{c} \.I. \"Unl\"ut\"urk}
\email{kunluturk17@ku.edu.tr}
\affiliation{Department of Physics, Ko\c{c} University, \\
Rumelifeneri Yolu, 34450 Sariyer, Istanbul, Turkey}

\author{Andrew Coates}
\email{andrew.coates.grav@gmail.com}
\affiliation{Department of Physics, Ko\c{c} University, \\
Rumelifeneri Yolu, 34450 Sariyer, Istanbul, Turkey}

\author{Fethi M. Ramazano\u{g}lu}
\email{framazanoglu@ku.edu.tr}
\affiliation{Department of Physics, Ko\c{c} University, \\
Rumelifeneri Yolu, 34450 Sariyer, Istanbul, Turkey}

\date{\today}
\maketitle

\section{Introduction}

Extensions of the massive vector field theory of Proca~\cite{Proca:1936fbw} with self-interactions find various applications in physics~\cite{Heisenberg:1936nmg, Conlon:2017hhi, Fukuda:2019ewf, dEnterria:2013zqi,Esposito-Farese:2009wbc, Annulli:2019fzq, Barton:2021wfj, Minamitsuji:2018kof, Herdeiro:2020jzx, Herdeiro:2021lwl}. Consequently, there have also been efforts to systematically categorize all such theories that are free of problematic degrees of freedom such as ghosts~\cite{tasinato2014cosmic, Heisenberg:2014rta, Heisenberg:2016eld, Kimura:2016rzw, Allys:2015sht}.

However, recently a body of results has shown that even the simplest self-interaction terms cause vector fields to have unphysical aspects~\cite{Clough:2022ygm, Mou:2022hqb, Coates:2022qia, Coates:2022nif,Coates:2023dmz}.\footnote{Also see \textcite{Esposito-Farese:2009wbc} for an earlier study.} Namely, the field can evolve for a finite amount of time after which the evolution becomes impossible due to a phenomenon called loss of hyperbolicity. The underlying reason is that the time evolution of the vector field is controlled by an effective metric, not the spacetime metric, which depends on the field values. Even when the spacetime metric is everywhere Lorentzian, the effective metric can change its signature at finite values of the vector field. Thus, the field equations become elliptic instead of hyperbolic.

So far, the existing literature has demonstrated the loss of hyperbolicity for the the specific case of a self-interaction potential where the mass potential $m^2 X_\mu X^\mu$ of the original Proca Lagrangian for a vector field $X^\mu$ is simply replaced with a self-interaction potential $V(X_\mu X^\mu)$. However, generalized Proca theories can contain more complicated terms such as derivative couplings. Even though the problems discovered in the case of the self-interaction potential have been predicted to be generic for any self-interaction, there has been no concrete study so far. Here, we address this issue, and show that vector fields with derivative self-couplings also suffer from a breakdown in their time evolution.

We examine the hyperbolicity of arguably the simplest generalized Proca theory with a derivative self-coupling, using techniques developed for the analysis of the self-interaction potential case~\cite{Clough:2022ygm,Coates:2022qia}. We numerically evolve initially healthy configurations, i.e., configurations for which the field equations are hyperbolic, and show that the equations can become elliptic after a finite time. While the overall problem is similar to those in the recent studies we mentioned, the derivative interaction also brings some novel ways the theory can break down.

We also show that beside loss of hyperbolicity, derivative self-interactions can lead to tachyonic instabilities for certain coupling functions and constants. This is a surprising result since tachyons are typically associated with interaction potentials rather than derivative couplings. The time evolution problem in this case is distinct from loss of hyperbolicity, nevertheless, it still renders the specific generalized Proca theory it appears in unphysical due to eternal exponential growth as we will explain.

\section{Generalized Proca theories}

\emph{Generalized Proca theories} aim to extend the Proca theory~\cite{Proca:1936fbw} to its most general form while avoiding problematic degrees of freedom, e.g. ghosts. All the allowed terms in $3+1$ dimensions lead to the Lagrangian~\cite{tasinato2014cosmic, Heisenberg:2014rta}
\begin{equation}
    \mathcal{L}_{\text{gen. Proca}} = -\frac{1}{4} F_{\mu\nu}F^{\mu\nu}
    + \sum_{n=2}^5 \alpha_n \mathcal{L}_n,
    \label{eq: generalized proca total lagrangian}
\end{equation}
where $F_{\mu\nu} = \nabla_\mu X_\nu - \nabla_\nu X_\mu$, and
\begin{subequations}
\begin{align}
    \mathcal{L}_2 &= f_2 (X_\mu, F_{\mu\nu}, \tilde{F}_{\mu\nu}) \label{eq: L2 subeq}\\
    \mathcal{L}_3 &= f_3(X^2) \nabla_\mu X^\mu
    \label{eq: L3 subeq}\\
    \mathcal{L}_4 &= f_4(X^2) \Big[ \left(\nabla_\mu X^\mu\right)^2 + c_2 \nabla_\rho X_\sigma \nabla^\rho X^\sigma \nonumber\\
    & \quad\quad\quad\quad\quad\quad\quad - \left(1+c_2\right) \nabla_\rho X_\sigma \nabla^\sigma X^\rho \Big]
    \label{eq: L4 subeq}\\
    \mathcal{L}_5 &= f_5(X^2) \Big[ \left(\nabla_\mu X^\mu\right)^3
    - 3 d_2 \nabla_\mu X^\mu \nabla_\rho X_\sigma \nabla^\rho X^\sigma \nonumber\\
    &\quad\quad\quad\quad\quad -3 \left(1-d_2\right) \nabla_\mu X^\mu \nabla_\rho X_\sigma \nabla^\sigma X^\rho \nonumber\\
    &\quad\quad\quad\quad\quad\quad + \left(2-3d_2\right) \nabla_\rho X_\sigma \nabla^\gamma X^\rho \nabla^\sigma X_\gamma \nonumber\\
    &\quad\quad\quad\quad\quad\quad\quad\quad +3d_2 \nabla_\rho X_\sigma \nabla^\gamma X^\rho \nabla_\gamma X^\sigma \Big].
\end{align}
\end{subequations}
$f_{2,3,4,5}$ are functions, $X^2=X_\mu X^\mu$, $\tilde{F}_{\mu\nu}$ is the dual of $F_{\mu\nu}$, and $\alpha_n$, $c_2$ and $d_2$ are constants. For example, the original Proca theory of a vector field with mass $m$ is simply $f_2 = -m^2 X^2/2$, with all other $f_i$ vanishing. Note that here and for the rest of the paper, we use the ``mostly plus" metric signature $(-,+,\dots, +)$ and $c=1$.

Generalized Proca theories have found applications to astrophysical systems~\cite{Chagoya_2016, PhysRevD.94.084039, PhysRevD.94.104039, Chagoya_2017, Heisenberg:2017hwb, minamitsuji2017black, PhysRevD.96.084049, Kase:2017egk, PhysRevD.95.123540, PhysRevD.96.084005, KASE2018541, PhysRevD.99.024052, Kase_2018, PhysRevD.102.024067, PhysRevD.105.044050} and cosmology~\cite{Jose_Beltran_Jimenez_2013, Emami_2017, DeFelice:2016yws, PhysRevD.94.044024, Heisenberg_2016, PhysRevD.95.104001, PhysRevD.99.063533, Kase_2018b, Domenech:2018vqj, doi:10.1142/S0218271819500640, Felice_2020, PhysRevD.102.124017, Heisenberg_2021, https://doi.org/10.1002/andp.202100453, GENG2021100819} and it has also been shown that they can lead to novel dynamics regarding screening mechanisms and coupling to alternative theories of gravity~\cite{DeFelice:2016cri,Heisenberg:2018acv, garcia2021coupling, Chagoya_2023, Ramazanoglu:2017xbl,Ramazanoglu:2019jrr}.

\subsection{Problems of the self-interaction potential}
\label{sec: self interaction potential}

Even though the above Lagrangian is designed to avoid ghosts, it has been recently shown that the generalized Proca theory with the simplest self-interaction potential beyond the mass term gives rise to ill-posed equations of motion. Namely, even though time evolution is possible, it cannot be continued beyond a certain point for some initial data, which is called loss of hyperbolicity. We will first summarize these results following the approach of \textcite{Coates:2022qia}, before we examine the case of derivative couplings.

The simplest generalized Proca theory is given by $\alpha_2=-1,\alpha_{i \neq 2}=0$ and
\begin{equation}
    f_2 = \frac12 m^2 X^2 + \frac{\lambda}{4} m^2 \left(X^2\right)^2 \equiv V(X^2) .
    \label{eq:self_interaction_potential}
\end{equation}
We shall call $V(X^2)$ the \emph{self-interaction potential}. Here, the constant $m$ is the mass parameter and $\lambda$ is a coupling constant that determines the strength of the self-interaction.

The field equations corresponding to Eq.~\eqref{eq:self_interaction_potential} are
\begin{equation}
    \nabla_\mu F^{\mu\nu} = 2V'X^\nu = m^2 \left(1+\lambda X^2\right) X^\nu,
\label{eq: sip field equation}
\end{equation}
where $V'=\mathrm{d}V/\mathrm{d}(X^2)$. It is not straightforward to deduce the well-posedness of Eq.~(\ref{eq: sip field equation}). In particular, the principal part, the term with the highest number of derivatives, of (\ref{eq: sip field equation}) is degenerate, in the sense that not the second ``time" derivative of each component of $X^\mu$ is present. One can, however, put Eq.~(\ref{eq: sip field equation}) into a form where answering these questions is easier. 

Note that just as in the free Proca case, we can take the divergence of (\ref{eq: sip field equation}) to get a generalized ``Lorenz condition"\footnote{We should emphasize that the Lorenz condition is a necessary condition that has to be satisfied by the massive vector field, it is not a gauge choice. There is no gauge freedom in this theory, unlike the case of Maxwell's theory.} which lets us express the divergence of $X^\mu$ in a different way, in terms of $X^\mu$ and its first derivatives:
\begin{equation}
    \nabla_\nu \left(z X^\nu\right) = 0 \quad \implies \quad \nabla_\nu X^\nu = - X^\nu \nabla_\nu \ln z.
    \label{eq: sip lorenz condition}
\end{equation}
Here, $z=1+\lambda X^2$ and we have used the antisymmetry of $F_{\mu\nu}$ to get $\nabla_\nu \nabla_\mu F^{\mu\nu} = 0$.
Using (\ref{eq: sip lorenz condition}), and the identity $\nabla_{\mu}F^{\mu\nu} = \square X^{\nu}-\nabla^{\nu}\nabla_\mu X^\mu - R^{\mu\nu}X_{\mu}$, where $R_{\mu\nu}$ is the Ricci tensor, one can write Eq.~(\ref{eq: sip field equation}) as
\begin{equation}
    \square X^\nu + \nabla^\nu \left(X^\mu \nabla_\mu \ln z \right) - R^{\mu\nu} X_\mu = m^2 z X^\nu,
\label{eq: sip manifestly hyperbolic form}
\end{equation}
with $\square=\nabla_\mu \nabla^\mu$. We can now analyze the hyperbolicity of (\ref{eq: sip manifestly hyperbolic form}), since the principal part reads as
\begin{equation}
    \square X^\nu + \frac{2\lambda}{z} X^\mu X_\alpha \nabla^\nu \nabla_\mu X^\alpha,
\label{eq: sip principal part}
\end{equation}
which is nondegenerate.

The field equation (\ref{eq: sip manifestly hyperbolic form}) is nonlinear; however, one can expand the equation around an arbitrary background solution $X_\mu^{(0)}$ to linear order, and this linearized problem should be well posed if the original problem is~\cite{Sarbach:2012pr}. Expanding (\ref{eq: sip manifestly hyperbolic form}) gives
\begin{align}
    \square X^\nu + \frac{2\lambda}{z^{(0)}} X_\mu^{(0)} & X_\alpha^{(0)} \nabla^\nu \nabla^\mu X^\alpha \nonumber\\
    &+\left(\text{lower order terms}\right)=0.
\end{align}
On small enough scales the background fields can be taken to be roughly constant, and the corresponding “frozen-coefficients” problem has to be well posed for the original problem to be well posed~\cite{Sarbach:2012pr}. Then, looking at the problem in Fourier space amounts to replacing $\nabla^{\nu}\nabla^{\mu}X^{\alpha}\to-k^{\nu}k^{\mu}\chi^{\alpha}$
where $\chi^{\alpha}$ is the Fourier transform of $X^{\alpha}$. We drop the superscripts (0) for convenience and proceed with the notation used by \textcite{Kovacs:2020ywu}. Now, in Fourier space the principal part of the linearized, frozen-coefficients problem reads as
\begin{equation}
    -k^\mu k_\mu \chi_\nu - \frac{2\lambda}{z} X^\mu X_\alpha k_\nu k_\mu \chi^\alpha \equiv -\mathcal{P}(k)^\beta_{\phantom{\beta}\nu} \chi_\beta,
\end{equation}
which defines the \emph{principal symbol}
\begin{equation}
    \mathcal{P}(k)^\beta_{\phantom{\beta}\nu} = k^\mu k_\mu \delta^\beta_\nu + \frac{2\lambda}{z} X^\mu X^\beta k_\nu k_\mu.
\end{equation}

The dispersion relation is determined by the condition $\det \mathcal{P}(k) = 0$, which can be written in $d+1$ dimensions as
\begin{equation}
\det\mathcal{P}=\left(g_{\mu\nu}k^{\mu}k^{\nu}\right)^{d}\frac{\left(\bar{g}_{\alpha\beta}k^{\alpha}k^{\beta}\right)}{z}=0,
\end{equation}
where we have defined the \emph{effective metric}
\begin{equation}
    \bar{g}_{\alpha\beta} = z g_{\alpha\beta} + 2\lambda X_\alpha X_\beta.
\label{eq: sip effective metric}
\end{equation}
We see therefore that there are two distinct types of modes: the standard ones that solve $g_{\mu\nu}k^{\mu}k^{\nu}=0$ and those that solve $\bar{g}_{\alpha\beta} k^{\alpha}k^{\beta}=0$. 

The final piece of the analysis is that $\gbar$ can change its sign for finite values of $X_\mu$. In rough terms, this means that the frequency $\omega$ given by the dispersion relation $\bar{g}_{\alpha\beta} k^{\alpha}k^{\beta}=0$ becomes imaginary, so that $e^{-i\omega t} \sim e^{|\omega|t}$. Hence, the corresponding modes grow exponentially instead of oscillating. As a result of this, the solution gets contaminated by infinitely many high $k$ modes which grow arbitrarily large in an arbitrarily short amount of time. More rigorously, $\bar{g}_{\alpha\beta}$ changing signature means that the partial differential equation ceases to be wavelike, and time evolution is not well defined any more~\cite{Stewart:2002vd,Sarbach:2012pr}. In short, the modes governed by $\bar{g}_{\alpha\beta}$ can lead to loss of hyperbolicity, and the transition from a hyperbolic system to an elliptic one is indicated by $\gbar_{\alpha\beta}$ becoming singular.

The above analysis using the principal symbol provides a procedure for analyzing the well-posedness of the equations of motion in any dimension. However, we also want to remark that in $(1+1)$ dimensions the analysis is especially simple, since in this case the equation of motion has \emph{exactly} the form of a generalized wave equation where the wave operator is controlled by the effective metric (\ref{eq: sip effective metric}).

To see this, we write the principal part (\ref{eq: sip principal part}) as
\begin{equation}
    \frac{1}{z} \bar{g}_{\alpha\beta} \nabla^\alpha \nabla^\beta X^\nu + \frac{2\lambda}{z} \Theta^\nu
\end{equation}
with $\Theta^\nu = X_\alpha X_\beta \nabla^\alpha F^{\nu\beta}$, which can be derived straightforwardly using the definition of $F^{\mu\nu}$ and the symmetries of the Riemann tensor. In general, $\Theta^\nu$ contains the second derivatives of the vector field through $\nabla_\alpha F^{\nu\beta}$ and thus contributes to the principal part. However, it is not difficult to show that in any spacetime with (1+1) dimensions, $\Theta^\nu$ vanishes (see the supplemental material of \textcite{Coates:2022qia}). Thus, in this specific case, the principal part of the partial differential equation is determined exactly by the wave operator $(1/z)\bar{g}_{\alpha\beta} \nabla^\alpha \nabla^\beta$.

These results already show that there are configurations of the vector field $X_\mu$ where time evolution simply does not exist. \textcite{Coates:2022qia} further showed that the vector field can evolve in such a way that the effective metric is initially Lorentzian, but  changes signature and becomes Euclidean. This so-called \emph{dynamical} loss of hyperbolicity occurs quite readily, e.g. it does not require spacetime curvature or an external driving source to grow the vector field. Moreover, such breakdown can even be observed starting from arbitrarily small field amplitudes~\cite{Coates:2023dmz}.

\section{Derivative coupling and the effective metric}

We saw that even the simplest self-interaction which solely depends on the field values gives rise to ill-posed equations of motion. Here we will explore how generic this pathology is by extending the same analysis to interactions that also depend on the derivatives of the vector field.

Arguably the simplest generalized Proca theory with derivative coupling arises from the $\mathcal{L}_{3}$ term (\ref{eq: L3 subeq}), together with the standard kinetic and mass terms, which has the total Lagrangian 
\begin{equation}
\label{eq:derivative_action_hyperbolicity}
\mathcal{L} = -\frac{1}{4}F_{\mu\nu}F^{\mu\nu}-\frac{1}{2}m^{2}X^{2}-\frac{1}{2}\lambda X^{2}\nabla_{\nu}X^{\nu},
\end{equation}
where $f_3=X^2$ and $\lambda =-2\alpha_3$ is a constant determining the strength of the self-interaction. The even simpler case of a constant $f_3$ does not contribute to the equation of motion since it provides a total derivative term. This theory yields the field equation
\begin{equation}
\nabla_{\mu}F^{\mu\nu}+\lambda\left(X_{\mu}\nabla^{\nu}X^{\mu}-X^{\nu}\nabla_{\mu}X^{\mu}\right)=m^{2}X^{\nu}.\label{eq: field equations covariant}
\end{equation}

We can again take the divergence of (\ref{eq: field equations covariant})
to obtain a generalized Lorenz condition, which is more complicated
than the one for the free Proca theory or the self-interaction potential. Indeed, applying $\nabla_{\nu}$
to Eq.~(\ref{eq: field equations covariant}) and using the fact $\nabla_{\mu}\nabla_{\nu}F^{\mu\nu}=0$
we get, after rearranging,
\begin{equation}
X_{\mu}\square X^{\mu}-X^{\mu}\nabla_{\mu}\rho=\rho^{2} + \frac{m^{2}}{\lambda}\rho - \nabla_{\mu} X_{\nu} \nabla^{\mu} X^{\nu},
\label{eq: lorenz condition}
\end{equation}
where we denote $\rho=\nabla_{\mu}X^{\mu}$ for brevity.

\subsection{Derivation of the effective metric}

One way to analyze the well-posedness of Eq.~\eqref{eq: field equations covariant} would be to re-express $\rho=\nabla_\mu X^\mu$ so that we can write the principal part of the problem in a nondegenerate way, just as we did in Sec.~\ref{sec: self interaction potential}. However, the generalized Lorenz condition Eq.~(\ref{eq: lorenz condition}) does not offer a way of algebraically solving for $\rho$ in terms of the vector field and its first derivatives, unlike the theory in Sec.~\ref{sec: self interaction potential}. We therefore proceed in a different manner to derive a quadratic equation for $\rho$ as follows.

Firstly, contracting Eq.~(\ref{eq: field equations covariant}) with $X_{\nu}$ and using $\nabla_{\mu}F^{\mu\nu}=\square X^{\nu}-\nabla^{\nu}\rho - R^{\mu\nu}X_{\mu}$, we get
\begin{align}
X_{\mu}\square X^{\mu} &- X^{\mu}\nabla_{\mu}\rho - R_{\mu\nu}X^{\mu}X^{\nu} \nonumber\\
& \quad + \lambda \left(X_{\mu}X_{\nu}\nabla^{\mu}X^{\nu}-X^{2}\rho \right) = m^{2}X^{2}.\label{eq: auxilliary eqn}
\end{align}
We can then use Eq.~(\ref{eq: lorenz condition}) to eliminate the derivative of $\rho$, which yields
\begin{equation}
\rho^{2}+a\rho+b=0.\label{eq: quadratic eqn for f}
\end{equation}
Here, $a=m^{2}/\lambda-\lambda X^{2}$ and $b=\lambda X_{\mu}X_{\nu}\nabla^{\mu}X^{\nu}-\nabla_{\mu}X_{\nu}\nabla^{\mu}X^{\nu}-R_{\mu\nu}X^{\mu}X^{\nu}-m^{2}X^{2}$.
Differentiating (\ref{eq: quadratic eqn for f}) allows us to express
the derivative of $\rho$ in a different way:
\begin{equation}
\nabla_{\nu}\rho=-\frac{\rho\nabla_{\nu}a+\nabla_{\nu}b}{2\rho+a}.
\end{equation}
Finally, inserting this into Eq.~(\ref{eq: field equations covariant}),
the field equations can be written as
\begin{align}
\square X^{\nu} &+ \frac{\rho\nabla^{\nu}a+\nabla^{\nu}b}{2\rho+a}-R^{\mu\nu}X_{\mu} \nonumber\\
& \quad +\lambda\left(X_{\mu}\nabla^{\nu}X^{\mu}-X^{\nu}\nabla_{\mu}X^{\mu}\right)=m^{2}X^{\nu},\label{eq: manifestly hyperbolic form}
\end{align}
whose principal part is
\begin{equation}
\square X^{\nu}+\frac{\nabla^{\nu}b}{2\rho+a}\ \sim\ \square X^{\nu}+\frac{\lambda X_{\alpha}X_{\beta}-2\nabla_{\alpha}X_{\beta}}{2\rho+a}\nabla^{\nu}\nabla^{\alpha}X^{\beta}.
\label{eq: derivative_principal_part}
\end{equation}

It can be readily seen that the principal part, Eq.~\eqref{eq: derivative_principal_part}, is nondegenerate; it includes the second time derivative of each component
of $X_{\mu}$. We can therefore proceed exactly as in Sec.~\ref{sec: self interaction potential} to analyze the well-posedness of the problem. We also note that Eqs.~(\ref{eq: quadratic eqn for f}) and (\ref{eq: manifestly hyperbolic form}) together are equivalent to the original field equation~(\ref{eq: field equations covariant}).

In Fourier space, the linearized, frozen-coefficients version of the principal part (\ref{eq: derivative_principal_part}) reads as
\begin{equation}
-k^{\alpha}k_{\alpha}\chi_{\nu}-\frac{\lambda X_{\alpha}X_{\beta}-2\nabla_{\alpha}X_{\beta}}{2\rho+a}k_{\nu}k^{\alpha}\chi^{\beta}\equiv-\mathcal{P}(k)_{\phantom{\nu}\nu}^{\beta}\chi_{\beta},
\end{equation}
from which we extract the principal symbol $\mathcal{P}(k)$:
\begin{equation}
\mathcal{P}(k)_{\phantom{\nu}\nu}^{\beta}=k^{\alpha}k_{\alpha}\delta_{\nu}^{\beta}+\frac{\lambda X_{\alpha}X^{\beta}-2\nabla_{\alpha}X^{\beta}}{2\rho+a}k_{\nu}k^{\alpha}.
\end{equation}

The dispersion relation is determined through $\det\mathcal{P}(k)=0$, which can be written in $d+1$ dimensions as
\begin{equation}
\det\mathcal{P}=\left(g_{\mu\nu}k^{\mu}k^{\nu}\right)^{d}\frac{\left(\bar{g}_{\alpha\beta}k^{\alpha}k^{\beta}\right)}{\left(2\rho+a\right)}=0,
\end{equation}
with the effective metric
\begin{equation}
\bar{g}_{\alpha\beta}=\left(2\rho+a\right)g_{\alpha\beta}+\lambda X_{\alpha}X_{\beta}-2\nabla_{(\alpha}X_{\beta)}.
\label{eq: effective metric definition}
\end{equation}
We therefore have the completely analogous result: aside from the standard modes that solve $g_{\mu\nu}k^{\mu}k^{\nu}=0$ we also have those that solve $\bar{g}_{\alpha\beta} k^{\alpha}k^{\beta}=0$. The modes governed by $\bar{g}_{\alpha\beta}$ can lead to a loss of hyperbolicity. For example, when the field values $X_\mu$ or their derivatives become large enough to dominate over the spacetime metric $g_{\mu\nu}$, there can be a change of signature in the effective metric $\gbar_{\mu\nu}$, which would directly indicate ill-posedness as in the case of the self-interaction potential in Sec.~\ref{sec: self interaction potential}.

We remark that, just as in Sec.~\ref{sec: self interaction potential}, in (1+1) dimensions the field equations have exactly the form of a generalized wave equation, where the wave operator is controlled by the effective metric. To see this, note that scaling the whole equation by $2\rho+a$ and discarding the lower order terms, the principal part \eqref{eq: derivative_principal_part} can be written as
\begin{equation}
    \bar{g}_{\alpha\beta}\nabla^\alpha \nabla^\beta X^\nu + \left(\lambda X_\alpha X_\beta - 2\nabla_\alpha X_\beta \right)\nabla^\alpha F^{\nu\beta} .
\end{equation}
In general, the term with $\nabla^\alpha F^{\nu\beta}$ also contributes to the principal part. However, in (1+1) dimensions $\nabla^\alpha F^{\nu\beta}$ is actually to be discarded from the principal part even if it is nonzero, for the following reason: In (1+1) dimensions, $\nabla_\alpha F^{\nu\beta}$ and $\nabla_\mu F^{\mu\nu}$ have exactly the same number of independent degrees of freedom, and so one can be entirely expressed in terms of the other. This means that as long as the field equations have the form $\nabla_\mu F^{\mu\nu} = y^\nu(X_\alpha, \nabla_\alpha X_\beta)$ with some functions $y^\nu$, as is the case here, $\nabla_\alpha F^{\nu\beta}$ can be re-expressed in terms of lower order derivatives. Thus, it does not contribute to the principal part, which is exactly the wave operator $\bar{g}_{\alpha\beta} \nabla^\alpha \nabla^\beta$.

\subsection{Time evolution and its breakdown}
\label{sec:time_breakdown}

We have already established that the dynamics of the vector field is not solely governed by the spacetime metric. In this section, we will first show that there are indeed configurations of $X_\mu$ for which time evolution is not possible. Even after this, one can argue that the theory of Eq.~(\ref{eq:derivative_action_hyperbolicity}) can still be redeemed if any initial data that can be time evolved stay so indefinitely. We will also show that this is not the case, that is, hyperbolicity can be lost dynamically when the effective metric starts as Lorentzian but evolves into a Euclidean one.

We can illustrate the breakdown of the theory in the simple case of the $(1+1)$-dimensional flat space using the Cartesian coordinates $(t,x)$. Introducing $(X^{t},X^{x})=(\phi,A)$ and $E=-\dot{A}-\partial_{x}\phi$, where the dot denotes a derivative with respect to $t$, Eqs. (\ref{eq: field equations covariant}) and (\ref{eq: lorenz condition}) can be written as the first order system
\begin{subequations}
\label{eq:time_evolution}
\begin{align}
\dot{A} & =-E-\partial_{x}\phi \label{eq: A time evolution}\\
\bar{g}_{tt}\dot{\phi} & =-\left(\phi\partial_{x}E+m^{2}A^{2}+\lambda A\phi\partial_{x}\phi\right)+\frac{m^{2}}{\lambda}\partial_{x}A \nonumber\\
& \quad\quad\quad +\dot{A}^{2}+\left(\partial_{x}\phi\right)^{2} \label{eq: phi time evolution}\\
\dot{E} & =m^{2}A+\lambda\left(\phi\partial_{x}\phi+A\dot{\phi}\right), \label{eq: E time evolution}
\end{align}
\end{subequations}
subject to the constraint
\begin{equation}
\mathcal{C}\equiv\partial_{x}E-\lambda A\left(E+\partial_{x}\phi\right)+\lambda\phi\partial_{x}A+m^{2}\phi=0,\label{eq: constraint}
\end{equation}
and the $tt$ component of the effective metric reads as
\begin{equation}
\bar{g}_{tt}=\lambda A^{2}-2\partial_{x}A-\frac{m^{2}}{\lambda}.
\end{equation}
Note that in $(1+1)$ dimensions the field $X_{\mu}$ has only one propagating degree of freedom, which means that the initial configuration of the two of $\left\{ \phi,A,E\right\} $ can be specified freely, while the third is to be computed from the constraint Eq. (\ref{eq: constraint}).

The determinant of the effective metric can be written as
\begin{equation}
    \bar{g} = E^2 + m^2 \left(\phi^2 - A^2 \right) - \left(\frac{m^2}{\lambda}\right)^2 - 2\phi\mathcal{C},
\end{equation}
which we will use to determine where, or if, the effective metric becomes singular. In order to see which initial data might evolve and later lead to loss of hyperbolicity, let us first investigate some configurations of the vector field that are outright problematic, i.e. have no well-posed time evolution to begin with.

A simple problematic configuration is when $X_{\mu}=0$ at a point, for which the determinant of the effective metric reads as $\left.\bar{g}\right|_{\phi=A=0}=E^{2}-(m^{2}/\lambda)^{2}$. It is therefore clear that no matter how small $\phi$ and $A$ initially are, the effective metric becomes positive definite, i.e., not Lorentzian, at some part of the space if the magnitude of $E=-\dot{A}-\partial_{x}\phi$ is sufficiently large there, i.e., the equations of motion are not hyperbolic. A simple example of such a configuration is given by
\begin{equation}
    A(0,x)=0,\quad E(0,x)=\alpha e^{-x^{2}/2\sigma^{2}},
    \label{eq: initial configuration}
\end{equation}
for which the constraint equation implies $\phi=-\partial_x E / m^2$ initially. Here, $\phi$ can be made arbitrarily small by increasing $\sigma$, and the determinant at the origin reads as $\left.\bar{g}\right|_{x=0}=\alpha^2 - (m^{2}/\lambda)^{2}$, which can be made positive by increasing the magnitude of $\alpha$.

We thus see that, since the effective metric depends not only on the field values, but on the derivatives of the field as well, even those configurations with arbitrarily small $X_{\mu}$ can be unhealthy to begin with if they include large enough $k$ modes in Fourier space.

In light of the above, to construct an initial configuration that evolves toward loss of hyperbolicity, let us concentrate on initial data with $A(0,x)=0$. Then, using the constraint equation (\ref{eq: constraint}), the time derivative of the determinant reads as
\begin{align}  
    \dot{\bar{g}} = \frac{2\lambda}{m^6} \partial_x E &\bigg\{ 2\left(\partial_x^2 E\right)^2 + m^2\Big[(\partial_x E)^2 \nonumber\\
    &\quad\quad\quad\quad\quad\quad\quad  + m^2 E^2 - E \partial_x^2 E \Big] \bigg\}.
\end{align}
Thus, as long as $E \partial_x^2 E < 0$, with a large enough $\lambda \partial_x E > 0$, the evolution seems to bring us toward $\bar{g}=0$ even if we start with a Lorentzian effective metric with $\bar{g}<0$. In the next subsection, we will explicitly confirm this expectation using numerical computations.

\subsection{Numerical time evolution and dynamical loss of hyperbolicity}

A definitive confirmation of the breakdown requires the fully nonlinear evolution of the vector field, which requires numerical methods. We closely follow the setup of \textcite{Coates:2022qia} and \textcite{Clough:2022ygm}, which we adapt for Eq.~\eqref{eq:time_evolution}. Namely, we integrate the PDE system using the method of lines with RK4 and fourth order spatial derivatives with fixed step sizes. We observe that no constraint damping or numerical dissipation is necessary, i.e. turning these off did not affect our solutions within numerical errors. All computations use a Courant-Friedrichs-Lewy factor of $\Delta t/\Delta x = 2^{-5}$. We explicitly test the convergence of our solutions as we present below.

Note that one can set $|m^2|=1$ and $|\lambda|=1$ by scaling the coordinates and the vector field values, unless these parameters vanish. Moreover, the field equations (\ref{eq: field equations covariant}) are invariant under the transformation
\begin{equation}
    \lambda \to -\lambda,\quad X_\mu \to -X_\mu,
    \label{eq: negative lambda transformation}
\end{equation}
so the solutions for positive and negative $\lambda$ are in one-to-one correspondence. In particular, under the transformation (\ref{eq: negative lambda transformation}), the effective metric (\ref{eq: effective metric definition}) is invariant up to a sign (which can be reabsorbed into its definition). Therefore, the existence of configurations that break down hyperbolicity is independent of the sign of the coupling constant $\lambda$. Our numerical results are for $m^2=1$, $\lambda=1$, but the conclusions are applicable to any value of $m^2>0$ and $\lambda$ due to the scaling properties we mentioned.

We consider the simplest setting of a $(1+1)$-dimensional flat Minkowski spacetime for our numerical example. We use the trivial foliation corresponding to the normal vector $n^{\mu}=(\partial_t)^\mu$ to the spatial slices, i.e. we use the familiar setup from Sec.~\ref{sec:time_breakdown}.

A possible numerical hindrance is the evolution toward $\bar{g}_{tt}=0$. Although this is merely a coordinate singularity and does not imply a physical breakdown of the theory~\cite{Coates:2022nif}, it nonetheless makes further numerical time evolution impossible in these coordinates as is obvious in Eq.~\eqref{eq: phi time evolution}. We therefore take care to avoid it and find a configuration that reaches $\bar{g}=0$ without encountering the coordinate singularity $\bar{g}_{tt}=0$.

Inspired by our observations on the analytical properties of the time evolution in Sec.~\ref{sec:time_breakdown}, we numerically evolved the following initial data
\begin{equation}
    A(0,x)=0,\quad E(0,x) = \alpha \left(1 + \beta x \right) e^{-x^{2}/2\sigma^2},
\label{healthy configuration}
\end{equation}
with $\phi=-\partial_x E / m^2$ initially.
This configuration guarantees that $E \partial_x^2 E < 0$ near $x=0$, and we also include the odd term with nonzero $\beta$ to control and increase $\partial_x E$.

\begin{figure}
    \centering
    \includegraphics[width=\columnwidth]{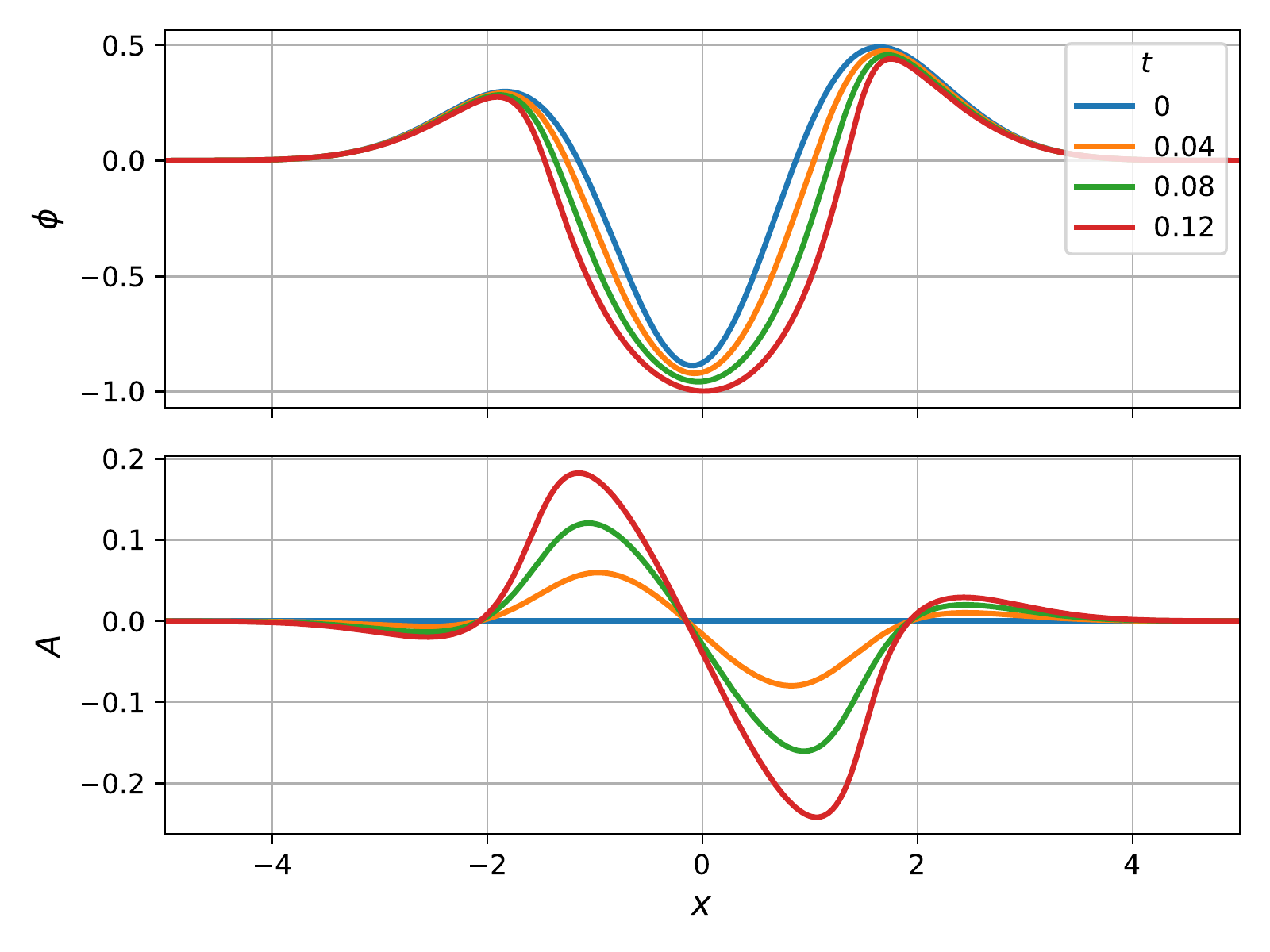}
    \caption{Evolution of the components of $X^\mu$ starting from the initial configuration (\ref{healthy configuration}). Here, $\alpha=0.25,\ \beta=3.5$ and $\sigma=m^2=\lambda=1$.}
    \label{fig: phi and A}
\end{figure}
\begin{figure}
    \centering
    \includegraphics[width=\columnwidth]{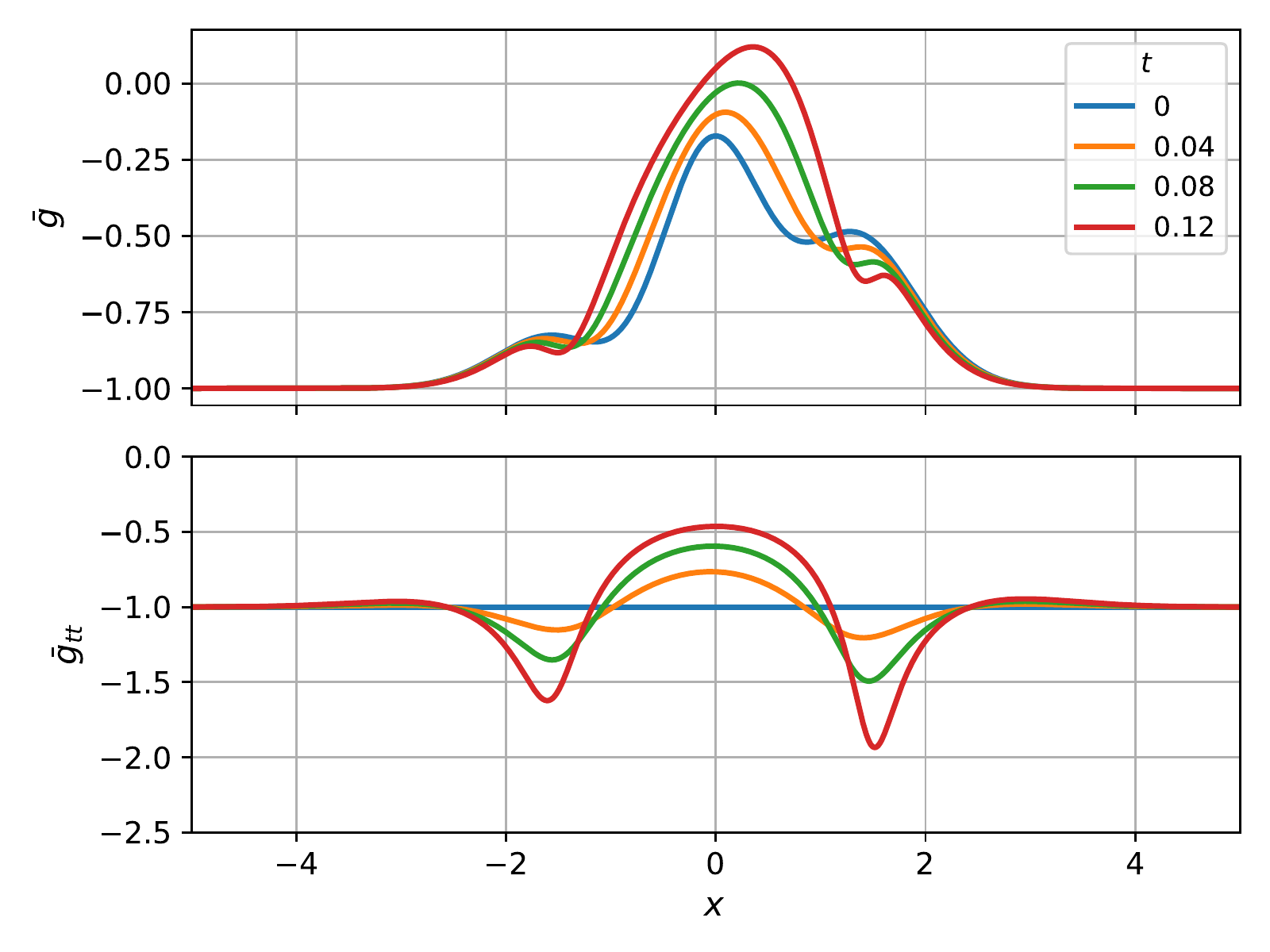}
    \caption{Dynamic loss of hyperbolicity during the evolution shown in Fig.~\ref{fig: phi and A}. An initially healthy ($\bar{g}<0$) configuration reaches the singularity and the equations of motion become elliptic ($\bar{g}>0$). The coordinate singularity $\bar{g}_{tt}=0$ is not encountered during the evolution.}
    \label{fig: gbar and gtt}
\end{figure}

In Fig.~\ref{fig: phi and A} we show the evolution of the $\phi$ and $A$ fields, and in Fig.~\ref{fig: gbar and gtt} we plot the determinant and $\gbar_{tt}$ for the same evolution. We can see that the vector field for which time evolution is initially possible evolves toward $\bar{g}=0$ and reaches this point, as expected. Hence, this shows that dynamical loss of hyperbolicity is not an isolated phenomenon for self-interaction potentials. On the contrary, the same time evolution problems occur quite readily in more complicated generalized Proca theories as well. Figure~\ref{fig: gbar and gtt} also explicitly shows that the evolution avoids encountering the coordinate singularity $\bar{g}_{tt}=0$.

\begin{figure}
    \centering
    \includegraphics[width=\columnwidth]{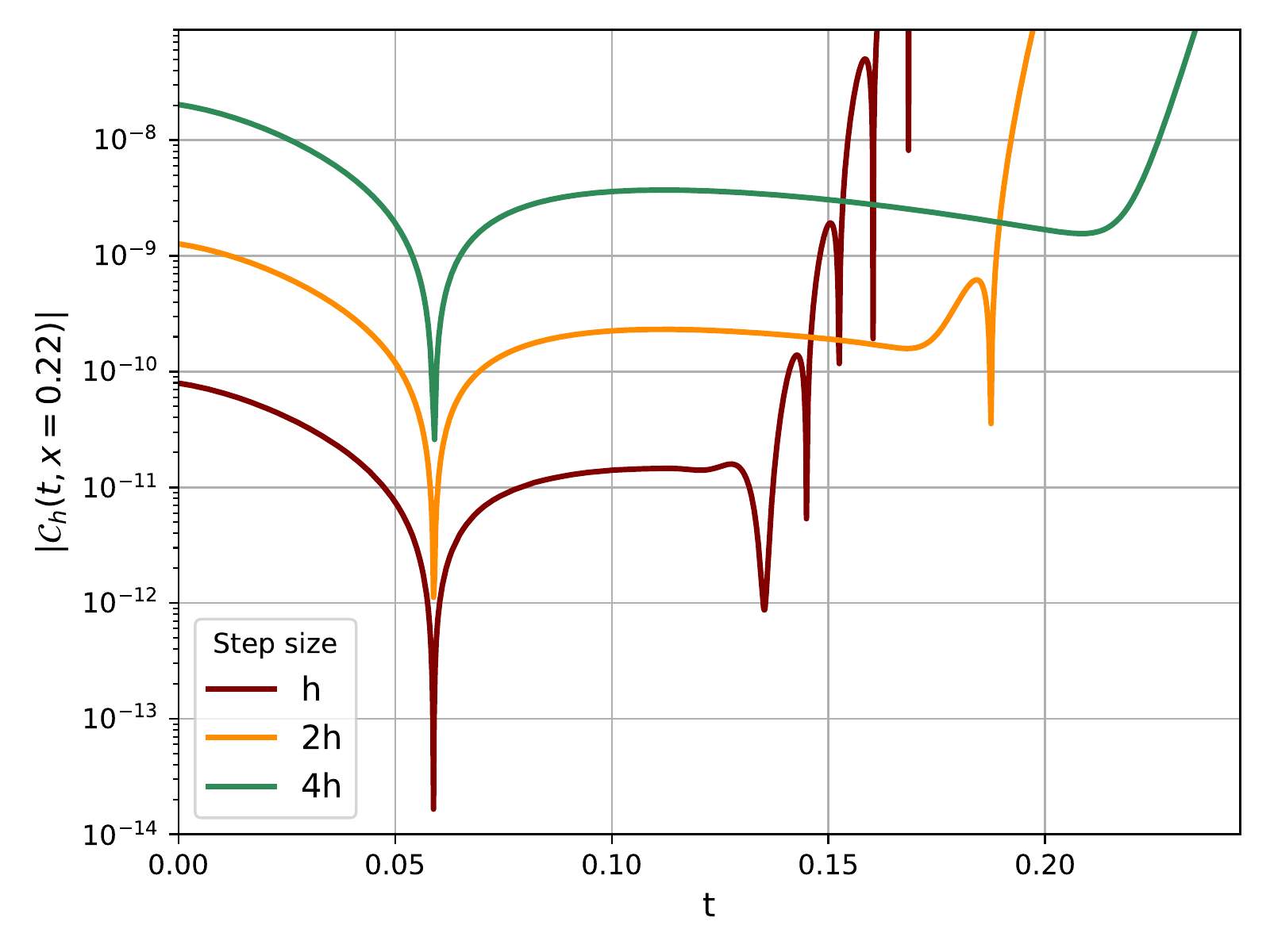}
    \caption{Time evolution of the constraint $\mathcal{C}$ at a fixed point of space, $x=0.22$, computed with different step sizes, where the base spatial step size is $\Delta x = h=2^{-8}$. The loss of hyperbolicity occurs at $t \approx 0.08$ at this spatial coordinate . As long as $\bar{g}<0$, finer resolutions provide smaller truncation errors and smaller constraints, i.e., the numerical solutions converge. However, when $\gbar \geq 0$ the trend is reversed, and there is divergence, since numerical solutions with smaller step sizes can accommodate (unphysical) higher frequency modes which grow faster. Exponential growth becomes noticeable at increasingly earlier times as $h$ decreases, as expected.
    } 
    \label{fig:divergence}
\end{figure}
It is important to recognize that time evolution is not well defined when $\gbar_{\alpha\beta}$ is not Lorentzian, that is, $\gbar \geq 0$, even though our numerical evolution continues in such regions in Figs.~\ref{fig: phi and A} and~\ref{fig: gbar and gtt}. This is due to the finite resolution of our numerics. In a truly continuous system, arbitrarily high frequency Fourier modes grow at arbitrarily fast exponential rates once $\gbar \geq 0$, and well-posed solutions cease to exist. In a finite difference scheme, there is a high frequency cutoff due to the nonzero spatial step size, which means the highest mode still has some high but finite growth rate (see, e.g., the supplemental material of \textcite{Coates:2022qia}). This behavior can be used to check loss of hyperbolicity. Whenever time evolution is healthy, finer resolutions provide smaller truncation errors, since the numerical solutions should converge to the exact solution in the limit of vanishing spatial step size. This behavior is reversed when $\gbar \geq 0$, since supposedly ``better'' numerical solutions with smaller step sizes can accommodate faster growing unphysical modes. Hence, there is divergence instead of convergence as the step size decreases, and the exponential growth also becomes visible earlier. We see all these characteristic features in Fig.~\ref{fig:divergence}.

\begin{figure}
    \centering
    \includegraphics[width=\columnwidth]{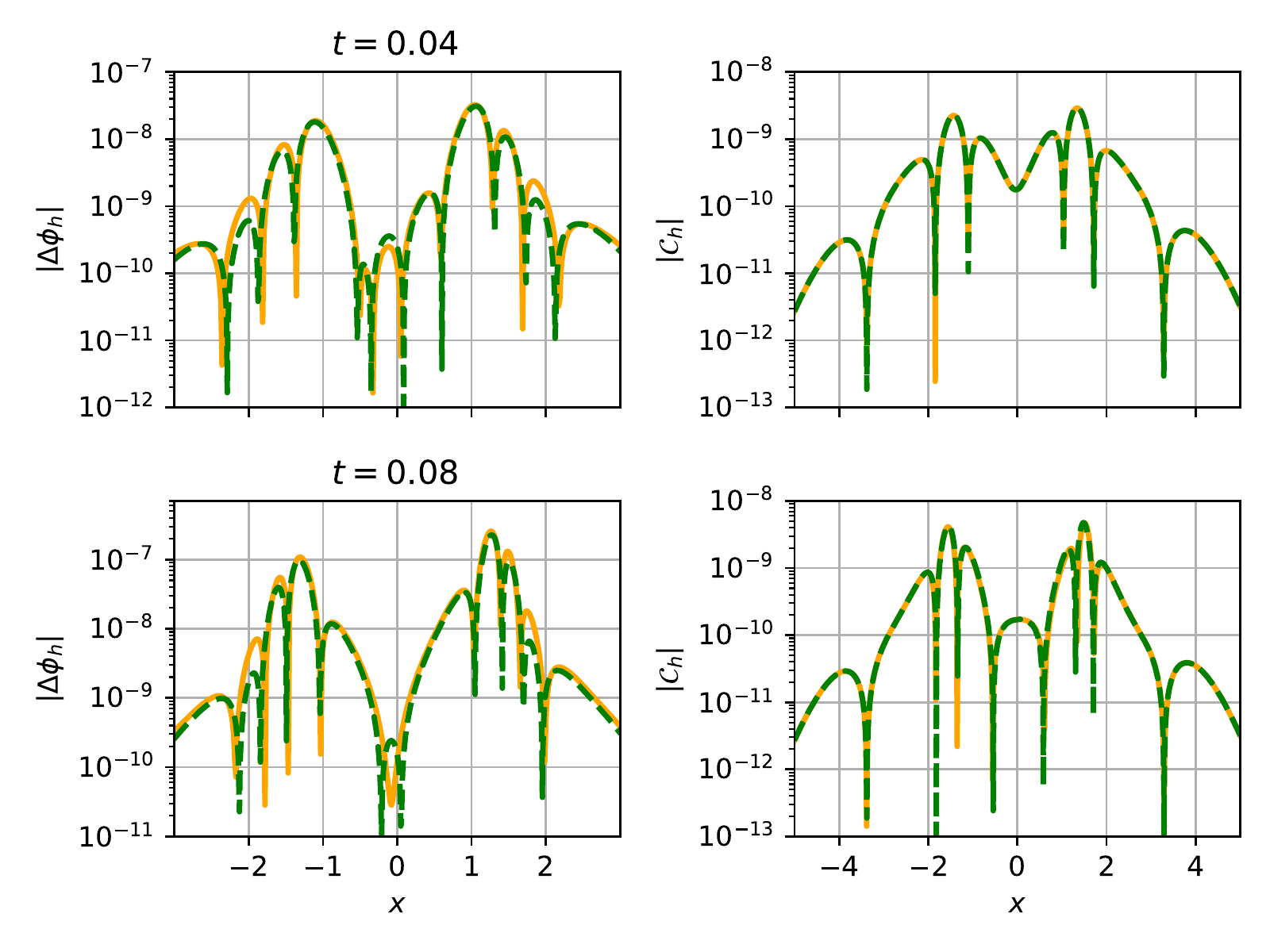}
    \caption{Fourth order convergence of the numerical solutions. Left: snapshots of the truncation error estimates $|\phi_{2h} - \phi_{4h}|$ (orange) and  $16|\phi_h - \phi_{2h}|$ (green) at two different times, where $\phi_h$ denotes the numerical value of $\phi$ calculated with a step size $h$. Right: the snapshots of $\mathcal{C}_{2h}$ (orange) and $16\mathcal{C}_{h}$ (green). Unlike $\phi$, the constraint itself, not merely the truncation error, converges to zero. The base step size is $h=2^{-8}$.}
    \label{fig: convergence}
\end{figure}
We repeat our numerical computations with different step sizes to further confirm that the results exhibit fourth order convergence whenever $\gbar<0$, as expected. The convergence of the constraint as well as the truncation error estimates for $\phi$ can be seen in Fig.~\ref{fig: convergence}.

\section{Derivative coupling and tachyons}
\label{ap: tachyonic instability}

The analysis of the time evolution of generalized Proca theories where all terms in Eq.~(\ref{eq: generalized proca total lagrangian}) are considered is quite involved, which we will not attempt here. However, we discovered another problematic behavior that arises from derivative couplings, which likely renders the theory it occurs in unphysical: tachyonic degrees of freedom.

Consider the Lagrangian term
\begin{align}
\mathcal{L}_{4} = \left(\nabla_{\mu}X^{\mu}\right)^{2} &+ c_{2}\nabla_{\rho}X_{\sigma}\nabla^{\rho}X^{\sigma} \nonumber\\
& \quad\quad -\left(1+c_{2}\right) \nabla_{\rho}X_{\sigma}\nabla^{\sigma}X^{\rho},
\end{align}
which corresponds to the simple choice $f_{4}(X^2)=1$ in Eq.~(\ref{eq: L4 subeq}). We can show that this term does not give rise to truly novel dynamics in a Ricci-flat background. That is, consider the action
\begin{equation}
S = \int\d^{4}x\sqrt{-g}\left(-\frac{1}{4}F_{\mu\nu}F^{\mu\nu}-\frac{1}{2}m^{2}X^{2}+\frac{1}{2}\alpha_{4}\mathcal{L}_{4}\right),
\label{eq: second order action}
\end{equation}
where again $X^{2}=X^{\mu}X_{\mu}$, $m$ is the mass parameter, and we have rescaled the coupling constant $\alpha_{4} \to \alpha_4 / 2$. The field equations are
\begin{align}\label{eq:tachyon_eom}
    \nabla_{\mu}F^{\mu\nu} - \alpha_{4}\big[&c_{2}\square X^{\nu}+\nabla^{\nu}\nabla_{\mu}X^{\mu} \nonumber\\ 
    & \quad - \left(1+c_{2}\right) \nabla_{\mu}\nabla^{\nu}X^{\mu} \big]=m^{2}X^{\nu}.
\end{align}

Using the Ricci tensor, these can be rewritten as
\begin{equation}
\left(1-\alpha_{4}c_{2}\right)\nabla_{\mu}F^{\mu\nu} + \alpha_4 R^{\mu\nu}X_\mu = m^{2}X^{\nu}.
\label{eq: linear eqn with ricci}
\end{equation}
On a Ricci-flat background ($R^{\mu\nu}=0$) this is exactly the field equation of a free Proca field, albeit with a renormalized mass $m_{\text{r}}^{2}=m^{2}/(1-\alpha_{4}c_{2})$. In particular, the field equation is hyperbolic on a Ricci-flat background.

However, it is manifest that Eq.~(\ref{eq: linear eqn with ricci}) may suffer from a different problem when $\alpha_4 c_2 > 1$. Let us first assume $R^{\mu\nu}=0$ for the sake of simplicity. Then, Eq.~(\ref{eq: linear eqn with ricci}) describes a wave equation for a vector field with a negative mass square, which is called a \emph{tachyon}. A tachyon leads to an infrared, i.e., low frequency, instability, since  $\omega^2 = \mathbf{k}^2 + m_{\text{r}}^2 < 0$ for low enough values of $\mathbf{k}$. As we have seen before, an imaginary frequency leads to the exponential growth of the corresponding Fourier mode, $e^{-i\omega t} \sim e^{|\omega|t}$. 

A tachyon is basically having the ``wrong'' sign in front of the mass potential term, and it is not necessarily problematic if  the initial growth can be quenched by some nonlinear terms. This is indeed the case for scalar field theories, where interaction potentials of the form $V(\varphi) = m^2 \phi^2+ \lambda^4 \phi^4$ are well behaved when $m^2<0$ and $\lambda^4>0$. The initial exponential growth due to the tachyon is eventually suppressed by the quartic term, which prevents an unbounded growth of the field. Indeed, such a controlled instability is the underlying mechanism of the Higgs mechanism~\cite{Kibble:2009} and the well-known spontaneous scalarization scenario in gravity~\cite{Damour:1993hw,Ramazanoglu:2016kul,Doneva:2022ewd}. 

The situation is quite different for a vector field theory however. First, the theory of Eq.~\eqref{eq:tachyon_eom} does not have any term other than the simple Proca mass; hence we have a pure tachyon that grows forever without bound. This would be a generic issue, and unless we consider an unnatural superposition of only high frequency modes, any initial data would blow up at a rate of $\sim e^{|m_{\text{r}}|t}$. The blowup likely persists when the Ricci tensor is not vanishing but sufficiently small. More importantly, adding a quartic self-interaction is not a viable option in this case, since we have already seen that this leads to loss of hyperbolicity.\footnote{This is closely related to the fact that generalizations of spontaneous scalarization to vector fields seem to be problematic, which itself is a recent discovery~\cite{Garcia-Saenz:2021uyv,Silva:2021jya,Demirboga:2021nrc,Doneva:2022ewd}.} Hence, Eq.~\eqref{eq:tachyon_eom} generically seems to lead to unphysical theories for a range of parameter values.

\section{Discussion}
It has already been shown that in the presence of the self-interaction potential the field equations of a vector field can dynamically lose hyperbolicity and become elliptic. Here we have explicitly shown that this problem is not unique to field self-couplings considered so far, but it is also present in the generalized Proca theories with derivative self-couplings. This confirms the expectation that loss of hyperbolicity is not an isolated phenomenon. Our concrete results are for simpler derivative couplings, but this further strengthens the already well-motivated expectation that generalized Proca theories suffer from similar problems in the generic case.

We examined the hyperbolicity of arguably the simplest generalized Proca theory that features a derivative self-coupling using the techniques developed by \textcite{Coates:2022qia}. That is, we formulate the field equation in such a way that its principal part is nondegenerate. This allows us to derive an effective metric which determines the character of the field equations. When the effective metric is Lorentzian (Euclidean), the field equations are hyperbolic (elliptic). Hence, when there is a signature transition, hyperbolicity is lost.

Since the effective metric depends on the vector field (and in our case its first derivatives as well), it may change signature as the vector field evolves, i.e. loss of hyperbolicity can occur dynamically. We have explicitly shown this by numerically evolving an initially healthy configuration where the effective metric eventually becomes Euclidean.

A significant difference of derivative self-couplings from the case of the self-interaction potential is the dependence of the effective metric on the derivatives of the vector field as well. As we have shown, this has the consequence that there exist very simple initial data that have arbitrarily small field amplitudes, for which which the equations of motion are elliptic to begin with, if the spatial derivatives are large enough. A single Fourier mode $e^{ikx}$ in the $k\to \infty$ limit is a simple example.

Our results establish the existence of configurations which break down in finite time; however, there also exist configurations which can evolve indefinitely without losing hyperbolicity. Nevertheless, we do not believe the configuration studied in this paper to be exceptional. For example, \textcite{Coates:2023dmz} showed that even configurations with arbitrarily small initial amplitudes can lose hyperbolicity in the case of the self-interaction potential. Although we have not shown this for the derivative couplings at hand, we expect similar configurations (e.g., infalling spherical waves) to break down.

We have also shown that beside loss of hyperbolicity, generalized Proca theories can suffer from tachyonic instabilities in some part of their parameter space, which leads to an unphysical eternal exponential growth of the fields. Unlike some examples in scalar field theories, there is no known mechanism to quench such a growth either. Overall, we expect this subset of the generalized Proca theories to be unphysical as well.

An important point to emphasize is that we have considered the actions we investigated so far to belong to fundamental theories; hence, the problems we discovered rendered them unphysical. However, it is always possible that such a theory constitutes an approximate description of a yet more fundamental theory. When we view generalized Proca theories as such \emph{effective field theories} (EFTs), these pathologies do not necessarily mean a breakdown of physics, but rather they may point to the breakdown of the approximation scheme we used in reducing the more fundamental theory~\cite{Barausse:2022rvg}. Even for such a view, however, the problems we identified have to be understood in detail if one wants to avoid computational problems in the study of these EFTs.

Problems about well-posedness in general and loss of hyperbolicity in particular have become an increasingly relevant topic of study in recent times, especially in gravity~\cite{Corman:2022xqg,R:2022hlf,Thaalba:2023fmq}. Analysis of these issues in the relatively simpler setting of vector fields enables us to better understand the underlying mechanisms. Moreover, identifying these pathologies can also help us in identifying alternative theories of gravity that are more likely to represent nature.

\acknowledgements
F. M. R. acknowledges support from T\"UB\.ITAK Project No. 122F097.

\end{document}